\documentclass{bmathcryptology}

\volume{1}
\issue{1}

\startpage{1}

\receiveddate{{\today}}
\revisiondate{{\today}}
\accepteddate{{\today}}

\articletitle{Quantum Equivalence of the DLP and CDHP for Group Actions}

\articleauthors{%
    Steven Galbraith\aff{1},
    Lorenz Panny\aff{2},
    Benjamin Smith\aff{3},
    Frederik Vercauteren\aff{4}
}

\articleaffiliations{%
    \aff{1}Mathematics Department, University of Auckland, New Zealand
    \\
    \aff{2}Department of Mathematics and Computer Science, Technische Universiteit Eindhoven, Netherlands
    \\
    \aff{\phantom{2}}Present affiliation: Institute of Information Science, Academia Sinica, Taipei, Taiwan
    \\
    \aff{3}Inria \emph{and} \'Ecole Polytechnique, Institut Polytechnique de Paris, Palaiseau, France
    \\
    \aff{4}imec-COSIC, ESAT, KU Leuven, Belgium
}

\correspondingauthoremail{s.galbraith@auckland.ac.nz, lorenz@yx7.cc, smith@lix.polytechnique.fr, frederik.vercauteren@kuleuven.be} 

\citationauthors{Galbraith, Panny, Smith, Vercauteren} 

\keywords{Quantum reduction, group action, discrete-logarithm problem, computational Diffie--Hellman problem} 

\MSClassification{%
    94A60,  
    68Q12,  
    11Y16   
  } 

\abstract{%
    In this short note we give a polynomial-time quantum reduction from the
    vectorization problem (DLP) to the parallelization problem (CDHP) for efficiently computable group actions.
    Combined with the trivial reduction from parallelization to vectorization,
    we thus prove the quantum equivalence of these problems,
    which is the post-quantum counterpart to classic results of den~Boer
    and Maurer in the classical Diffie--Hellman setting.  In contrast to the classical
    setting, our reduction holds unconditionally and does not assume knowledge of
    suitable auxiliary algebraic groups.
    We discuss the implications of this reduction
    for isogeny-based cryptosystems including CSIDH.
}

\usepackage{tikz}

\usepackage{xcolor}
\definecolor{linkcolor}{rgb}{0.65,0,0}
\definecolor{citecolor}{rgb}{0,0.65,0}
\definecolor{urlcolor}{rgb}{0,0,0.65}

\usepackage[british]{babel}

\newtheorem*{remark*}{Remark}{\normalfont\itshape}{\normalfont}

\newcommand{\F}{\mathbb{F}}

\newcommand{\OO}{\mathcal{O}}

\newcommand{\Z}{\mathbb{Z}}

\newcommand{\cl}{\operatorname{cl}}

\newcommand{\End}{\mathrm{End}}

\renewcommand{\a}{\mathfrak{a}}
\renewcommand{\b}{\mathfrak{b}}

\renewcommand{\l}{\mathfrak{l}}

\newcommand{\g}{\mathfrak{g}}
\newcommand{\x}{\mathfrak{x}}

\newcommand\act{{\hspace{.14em}{\ast}\hspace{.14em}}}

\newcommand\subheading[1]{\par\vspace{3ex}\noindent\textbf{#1}}
\newcommand\myqed{}

\usepackage{enumitem}
\setlist{topsep=+.9ex,itemsep=-.7ex}

\begin{document}

\begin{NoHyper}
\articleinformation 
\end{NoHyper}


\begingroup
  \makeatletter
  \def\@thefnmark{}\relax
  \@footnotetext{\relax
  \textsuperscript{*}%
  Author list in alphabetical order; see
  \url{https://www.ams.org/profession/leaders/culture/CultureStatement04.pdf}.
  This work was supported in part
  by the Ministry for Business, Innvovation and Employment in New Zealand project UOAX1933,
  by the Commission of the European Communities through the Horizon 2020 program under project numbers 643161 (ECRYPT-NET) and 830892 (SPARTA),
  by the French Agence Nationale de la Recherche through ANR CIAO (ANR-19-CE48-0008),
  by the Research Council KU Leuven grant C14/18/067,
  and by CyberSecurity Research Flanders with reference number VR20192203.
  }
\endgroup

\section{Introduction}
\noindent
In their seminal 1976 paper~\cite{DH76},
Diffie and Hellman conjectured
that breaking their new key exchange protocol
(in the sense of computing the shared secret from the public keys)
was as hard as computing discrete logarithms.
This polynomial-time equivalence was later proven
(assuming knowledge of suitable auxiliary algebraic groups
of smooth order)
for all groups by Maurer~\cite{Maurer},
based on earlier results of den~Boer~\cite{denBoer}
covering certain special cases.

In this short paper,
we prove an unconditional reduction
between the analogous problems for group actions
in the quantum setting.
This result has important implications for the quantum security
of the \emph{CSIDH} key-exchange scheme~\cite{csidh}.

\begin{figure}[h]
    \begin{center}
    \begin{tikzpicture}[
            every path/.append style={->,line width=.8pt,shorten <=1pt,shorten >=1pt,>=stealth},
            yscale=1.5,
        ]
        \definecolor{darkgreen}{RGB}{0,159,31}
        \begin{scope}[every node/.append style={circle,fill=black,inner sep=1.5pt}]
            \node (v0) at (0,0) {};
            \node (v1) at (2,1) {};
            \node (p0) at (4,0) {};
            \node (pl) at (6,1) {};
            \node (pr) at (6,-.2) {};
        \end{scope}
        \node (p1) at (8,.8) {};
        \draw[darkgreen] (v0) -- (v1);
        \draw[dotted,shorten <=2pt] (p0) -- (pl);
        \draw[dotted,shorten <=2pt] (p0) -- (pr);
        \draw[dotted,shorten >=2.8pt] (pl) -- (p1);
        \draw[dotted,shorten >=3.3pt] (pr) -- (p1);
        \draw (p1) node[darkgreen,rectangle,draw,inner sep=4pt] {};
    \end{tikzpicture}
    \end{center}
    \vspace{-2ex}
    \caption{
        The vectorization and parallelization problems.
    }
    \vspace{-1ex}
    \label{fig:vec_par}
\end{figure}
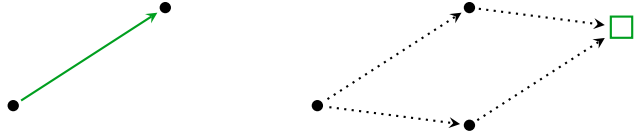

\subheading{Cryptographic group actions.}
In 1997, Couveignes introduced the notion of a
\emph{hard homogeneous space}~\cite{hhs},
essentially
a free and transitive finite abelian group action
$\act\colon G\times X\to X$
which is efficiently computable%
\footnote{See Section~\ref{sec:effgrpact} for a precise definition.}
while other computational problems are hard.
In Couveignes' terminology,
these
are \emph{vectorization}
and \emph{parallelization}, named by analogy with the archetypical
example of a homogeneous space: a vector space acting on
affine space by translations (cf.\ Figure~\ref{fig:vec_par}).
The vectorization problem is: given $x$ and $g\act x$ in $X$, compute $g\in G$.
The parallelization problem is: given $x$, $g\act x$, and $h\act x$ in $X$,
compute $gh\act x\in X$.
The group-exponentiation analogues of these problems
are the
\emph{discrete logarithm problem} (DLP)
and
\emph{computational Diffie--Hellman problem} (CDHP).

For twenty years, there was little interest in the
hard-homogeneous-spaces framework, since all known
(conjectural) instantiations were
either painfully slow in practice
or already captured by the group-exponentiation point of view.
However, interest in these one-way group actions has reemerged
due to the current focus on post-quantum cryptography,
where group-exponentiation Diffie--Hellman is broken in
polynomial time by Shor's algorithm~\cite{shor},
but group \emph{actions} are not.
In particular, \emph{CSIDH} is a
cryptographic
group action
that appears to be post-quantum secure
and reasonably efficient in many scenarios~\cite{csidh}.

\subheading{DLP--CDHP reductions.}
Just like in the classical group-exponentiation setting,
it is evident that parallelization
reduces to vectorization: recover $x$ from $x \act g$, then
apply $x$ to $y \act g$ to obtain $xy\act g$.
Traditionally,
the other direction is much more subtle.
The reduction essentially relies on the existence of auxiliary algebraic groups
of smooth group order over $\F_{q_i}$, where the $q_i$ are the prime divisors
of the order of the group in which the DLP and CDHP are defined.

The first result was given by den Boer~\cite{denBoer}, who showed the DLP and CDHP
to be equivalent in $\F_p^\times$ when $p$ is a prime such that the Euler totient $\varphi(p-1)$ is smooth.
The auxiliary groups are simply $\F_{q_i}^\times$ for each prime divisor $q_i \mid p-1$,
and the smoothness assumption implies that the DLP in each    $\F_{q_i}^\times$ is easy.
Maurer~\cite{Maurer} generalized this result to arbitrary cyclic groups $G$,
assuming that for each large prime divisor $q_i$ of $|G|$,
there exists an efficiently constructible elliptic curve $E/\F_{q_i}$ with smooth group order.

These reductions do not apply in the group-action setting
on classical computers~\cite[\S11]{smith}.
However, we show that there exists a polynomial-time
\emph{quantum} reduction from the vectorization to the
parallelization problem for group actions,
without relying on any extra assumptions.
This proves the polynomial-time equivalence of these problems in the quantum setting.

\section{Efficient group actions}\label{sec:effgrpact}
\noindent
We now define what it means for a group action $G\times X\to X$
to be ``efficiently computable''.
Since our main motivation is CSIDH (where $G$ is an ideal class group and $X$
is a set of elliptic curves), we use the notation  $\a,\b,\dots$
for elements of the group $G$, and denote by $E$ an element of the set $X$.

\begin{definition}
Let $G$ be a finite abelian group and $X$ a finite set.
    We abbreviate ``polynomial in $\log(\lvert G\rvert+\lvert X\rvert)$'' as ``polynomial''.
A group action
$\act\colon G\times X\to X$
is \emph{efficiently computable}
if
all elements of $G$ and $X$
have
(not necessarily unique)
bit representations
of polynomial length,
a generating set of~$G$ of polynomial size is given,
and
the following tasks can be
performed in polynomial time:
\begin{enumerate}
\item Compute the composition $\a\b \in G$ of any $\a,\b\in G$.
\item Compute the action $\a \act E$ of any $\a \in G$ on any $E \in X$.
\item Represent elements of $X$ canonically as bit strings. \label{item:bitrep}
\end{enumerate}

\noindent
    \emph{Vectorization} is:
    Given $E{\,\in\,} X$ and $E'{\,\in\,} G\act E$,
    compute any $\a{\,\in\,} G$ with $E'{\,=\,}\a\act E$.

\noindent
    \emph{Parallelization} is:
    Given $E{\,\in\,} X$ and ${\a\act E, \b\act E\in G\act E}$,
    compute $\a \b\act E {\,\in\,} X$.%
    \footnote{%
        The apparent ambiguity in
        the choice of
        $\a$ and $\b$
        lies in the
        stabilizer subgroup of $E$,
        thus cancels out
        in the result
        $\a \b\act E$.
    }
\end{definition}

\begin{remark*}
    The notion of a
    ``hard homogeneous space''
    as defined by Couveignes~\cite{hhs}
    additionally
    requires that~$\act$
    is free and transitive,
    that uniform sampling from~$G$ is polynomial-time,
    and that vectorization and parallelization are hard for~$\act$.
    On the other hand,
    Task~\ref{item:bitrep}
    is weakened to
    efficient
    membership and equality testing.
\end{remark*}

\section{The reduction}
\noindent
Let $\pi$ be an algorithm that solves the parallelization problem for
an efficient group action
$G\times X\to X$.
In other words, $\pi$
takes $\a\act E$ and $\b\act E$ and returns $\a\b\act E$.
We show that oracle access to a quantum circuit that computes $\pi$
allows one to solve the vectorization problem for \(\act \colon G\times X\to X\) in polynomial time.

\begin{lemma}
    \label{lemma:sqm}
    Given
    an element $\a\act E\in X$
    and
    access to a parallelization oracle~$\pi$,
    one can for any integer $n\geq0$
    compute $\a^n\act E$
    using $\Theta(\log n)$ queries to $\pi$.
\end{lemma}
\begin{proof}
    We perform double-and-add in the ``implicit group''~\cite{smith}
    of exponents, using the 
    oracle
    $\pi\colon(\a^x\act E,\,\a^y\act E)\mapsto \a^{x+y}\act E$
    for addition and doubling.
\myqed\end{proof}

\begin{theorem}
    \label{theorem:main}
 Let $\act \colon G\times X\to X$ be an efficiently computable group action.
    Given
    quantum access to
    a perfect
    parallelization oracle $\pi$,
    one can construct a quantum algorithm
    for the vectorization problem
    that runs in
    polynomial time.
\end{theorem}
\begin{proof}

We are
given an instance $(E,\a\act E)\in X^2$
of the vectorization problem.%
\footnote{%
    The element $\a$ is only defined up to $\mathrm{Stab}(E)$,
    but this choice will cancel throughout.
    }

\vspace{.4ex}

From the public description of $G$,
    we get a polynomially-sized generating set $\g_1,...,\g_r$.
    For $\underline x\in\Z^r$,
    write $\g^{\underline x}=\prod_{i=1}^s \g_i^{x_i}$,
    and define the map
    \begin{align*}
        h\colon\
        \Z^r\!\!
        &\ \longrightarrow\,X
        \\
        \underline x
        &\ \longmapsto\
        \g^{\underline x} \mathbin\act E
        \,\text.
    \end{align*}
    We apply
    Boneh and Lipton's~\cite{boneh-lipton}
    or Kitaev's~\cite{kitaev}
    higher-dimensional
    generalisation of Shor's algorithm~\cite{shor}
    to compute the period lattice
    \[
        K=\{\,\underline x\in \Z^r : \g^{\underline x}\act E=E\,\}
    \]
    of the map $h$
    in polynomial time.
    Note that $\Z^r/K$ is isomorphic to $G/\mathrm{Stab}(E)$.

\vspace{.4ex}

Now, define
\vspace{-2ex}
    \begin{align*}
        f\colon\
        \Z^r\times\Z
        &\ \longrightarrow\ X
        \\
        (\underline x, y)
        &\ \longmapsto\
        \g^{\underline x} \mathbin\act (\a^y \act E)
        \,\text.
    \end{align*}
Observe that
    $\a^y\act E$ can be computed using Lemma~\ref{lemma:sqm}:
    Negative $y$ may be replaced by a positive representative
    modulo $\det(K)$, which must be a multiple of the order of~$\a\cdot \mathrm{Stab}(E)$.
Thus,
using the efficient algorithm for the group action
and
the oracle access to~$\pi$,
one can
construct a quantum circuit that computes $f$ in polynomial time.
    The function $f$ is a homomorphism
    to the implicit group on the orbit of $E$ isomorphic to $\Z^r/K$,
    hence defines an instance of the hidden-subgroup problem
    with respect to its kernel, i.e., the lattice
    \[
        L = \{\, (\underline x,y)\in\Z^r\times\Z \,:\, \g^{\underline x + y \underline v} \act E = E \,\}
        \,\text,
    \]
    where $\underline v\in\Z^r$ is any vector such that $\g^{\underline v}\act E=\a\act E$.%
    \footnote{%
        Note that $\underline v$ is only defined modulo~$K$,
        but this does not matter since
        $L\supseteq K{\times}\{0\}$.
    }
    This (abelian) hidden-subgroup problem can be solved in polynomial time
    again using Shor's algorithm,
    making use of the efficient circuit to compute $f$
    constructed above.
    Finally, any vector in $L$ of the form $(\underline x,1)$
    satisfies $\g^{-\underline x}\act E=\a\act E$,
    hence yields a solution to the vectorization problem.
\myqed\end{proof}

\begin{remark*}
    If desired,
    the generating set $\g_1,...,\g_r$
    can be replaced by a smaller
    generating set after computing~$K$
    and before defining~$f$.
    Moreover,
    if elements of $G$ have
    unique representation,
    the computation of $K$
    can be replaced by a
    group-structure computation;
    the benefit is that this is
    independent of $E$,
    hence can be amortized across
    multiple vectorization instances.

    Also note that the computation
    of $K$ is only necessary to handle
    negative $y$ when evaluating $f$;
    hence,
    it seems this step could be omitted
    by using a variant of Shor's
    algorithm that only queries $f$ on
    the subset
    $\Z^r\times\Z_{\geq0}$.
    The computation of $K$ can also
    be skipped if the order of $G$ is known a priori,
    or if the action of inverses can be computed
    in a different way:
    For example,
    in the CSIDH setting,
    when $E$ is the starting curve
    chosen in~\cite{csidh},
    then $\x^{-1}\act E$ can be obtained
    as the quadratic twist of $\x\act E$.
\end{remark*}


\subsection{Imperfect oracles}
    It is unclear how to perform the reduction above when
    $\pi$ is only guaranteed to succeed with non-negligible probability $\alpha$,
    meaning that the probability over all triples $(E, \a\act E,\b\act E) \in X^3$
    that the oracle outputs $\a\b\act E$ is at least $\alpha$.

    In the classical discrete-logarithm setting, it is
    straightforward to amplify the success probability of CDH oracles
    using a random self-reduction of problem instances~\cite{MaWo96,shoup}:
    one computes lists of possible values of $g^{ab}$
    by blinding the inputs and unblinding the outputs,
    and uses majority vote to determine the correct result.
    Any exponentially small failure probability can be achieved
    using polynomially many queries~\cite[\S\,5]{shoup}.

    In the group-action setting, however, blinding does not work:
    The results cited above use a
    blinding map of the form $g^a\mapsto (g^a)^xg^y = g^{ax+y}$,
    which relies on the fact that we can multiply two public keys.
    But the best we can do for a mere group action is
    to translate the inputs by random elements,
    i.e., blind
    as $\a\act E\mapsto \x\act(\a\act E)$
    with a random $\x\in G$,
    which is insufficient:
    For example, if $\mathcal A$ is a perfect CDH oracle, then
    the oracle $\mathcal B$ that returns the output of $\mathcal A$
    either unmodified (with probability $\epsilon$),
    or shifted by a fixed element $\mathfrak z\in G$,
    is entirely unaffected by blinding and hence cannot be
    amplified using this idea.
    Thus, we must unfortunately leave the case of imperfect oracles
    as an open problem.

\section{Implications for CSIDH}\label{sec:csidh}
\noindent
Let $E$ be an elliptic curve over $\F_q$ with $\End_{\F_q}(E) = \OO$ being an order in an imaginary quadratic field.
Any invertible $\OO$-ideal $\a$ gives rise to an isogeny \(\varphi_\a\colon E \to E'\)
with kernel \(E[\a] = \{ P \in E( \overline{\F_q} ) : \forall \psi {\,\in\,} \a,\; \psi(P) = 0 \} \).
This leads to an action of $\cl(\OO)$ on a set $X$ of elliptic curves isogenous to $E$ and with the same endomorphism ring as $E$.
Precisely, $\a \act E := E' = E/E[\a]$.
This is the homogeneous space underlying CSIDH~\cite{csidh},  the Couveignes and Rostovtsev--Stolbunov
cryptosystems~\cite{hhs,Stolbunov10,DFKS},
and also the SeaSign signature scheme~\cite{SeaSign}.

Public keys are instances \((E,\a\act E)\) of the vectorization problem
in this homogeneous space.
In CSIDH, the Diffie--Hellman secret shared between Alice and Bob,
with public keys \((E, \a\act E)\) and \((E, \b\act E)\),
is \(\a\b\act E\).  Recovering the shared secret from the public keys is therefore solving a
parallelization problem.

Unfortunately, CSIDH is not
known to be
an efficiently computable group action in general.
The standard implementations of CSIDH~\cite{csidh} 
use secret keys of the form
$\a=\prod_i\l_i^{e_i}$,
where \(\underline e=(e_1,\dots,e_n)\in\Z^n\) are short exponent vectors
and the $\l_i$ are a fixed set of ``small'' ideals
whose action is efficient.
The action of~$\a$
is then evaluated as repeated applications of the $\l_i$ and their inverses.
However,
with these implementations,
it is no longer efficient
to
evaluate
the action of
a \emph{composition}
of such ``nice'' ideals:
a sequence of $k$
additions
starting from
short
exponent vectors
can result in an
exponent vector of
$1$-norm
exponential in $k$.

If one sets up a sequence of CSIDH instantiations for unbounded security levels, then there is no known polynomial-time method to sample uniformly from the groups $G$ or compute the action in polynomial time.
There are two reasons for this.
First, one might need relatively large prime ideals $\l_i$ to generate the class group.
Second, and more serious, given a randomly chosen ideal $\a$ it may be hard to find a short representation of an equivalent ideal of the form $\prod_i\l_i^{e_i}$.
Even when the class group structure is known, finding a short exponent vector \(\underline{e}\) requires solving a close(st)-vector problem for the relation lattice $\ker(\Z^n{\,\to\,}\cl(\OO))$.
But asymptotically,
polynomial-time lattice reduction algorithms
cannot guarantee that the output will have norm small enough
to ensure that the resulting group action is
computable in polynomial time.

For the above reasons, Theorem~\ref{theorem:main} does not apply directly to the general case of CSIDH or related cryptographic systems.
However, this does not mean the result has no practical meaning.
For example, since the dimensions $n$ used in CSIDH are rather small
(e.g.\ the \mbox{CSIDH-512} parameter set from~\cite{csidh} uses $n = 74$),
an efficient lattice-reduction algorithm such as BKZ~\cite{BKZ} with moderate block size suffices 
to obtain highly practical results (reducing a random relation lattice of dimension 74
using BKZ with block size 20 yields exponent vectors only 8 times longer than normal
\mbox{CSIDH-512} private keys).
As another example, the CSI-FiSh~\cite{CSI-FiSh} system has a known relation lattice and a relatively efficient group operation, so our theorem shows that the parallelization and vectorization problems are equivalent in a practical sense for this system.
Similarly, we expect that in many reasonable cryptographic settings
(possibly after some quantum and classical precomputation) our result
will provide a meaningful equivalence of the parallelization and vectorization problems.


\printbibliography 


\end{document}